\documentclass[12pt,a4paper]{article}
\usepackage{amssymb,amsmath}
\usepackage[T2A]{fontenc}
\usepackage[cp866]{inputenc}
\usepackage[russian,english]{babel}
\usepackage{hyperref}
\usepackage{graphicx}
\usepackage[dvips]{psfrag}
\newcommand{\risheight}{10cm}
\newcommand{\lambdabar}{{\mkern0.75mu\mathchar '26\mkern -9.75mu\lambda}}
\begin{document}
\begin{center}
{\Large The atomic nucleus as a bound system of $3A$ quarks}\\[3mm]
{B.~P.~Kosyakov${}^a$,
E.~Yu.~Popov${}^a$,
and  M. A. Vronski{\u\i}${}^{a,b}$
}\\[3mm]
{{\small 
${}^a$Russian Federal Nuclear Center--VNIIEF, Sarov, 607188 Nizhni{\u\i} Novgorod Region, Russia\\
${}^b$Sarov Institute of Physics {\&} Technology, Sarov, 607184 Nizhny Novgorod Region, Russia
}} 
\end{center}
\begin{abstract}
\noindent

The atomic nucleus, viewed as a system of bound quarks, should, in principle, be described
within an effective theory of  low-energy quantum chromodynamics. 
This paper provides an overview of recently developed models that embody essential features of the desired effective theory.
The Fermi gas model helps explain why the number of $d$ quarks is approximately equal to that of $u$ quarks  in stable light nuclei up to ${\rm {}^{40}_{20}Ca}$.
A modified bag model accounts for the deviation from this rule in heavier nuclei.
With this model, the static properties of a wide range of stable nuclei can be described with reasonable accuracy.
To make the most of the modified bag model, it is useful to invoke gauge/gravity duality.
A refined version of duality states: ``The dynamics inside an extremal black hole in ${\rm AdS}_5$ is mapped onto the corresponding dynamics of a stable subnuclear system in ${\mathbb R}_{1,3}$''.
This version of duality allows one to predict the primary decay channel of the lightest glueball.
Another implication is that this framework explains why the periodic table contains a finite number of stable elements.
Duality makes it possible to calculate the maximum allowed charge  $Z_{\rm max}$ of stable heavy nuclei: $Z_{\rm max}\approx 82$, which is the charge of the  ${\rm {}^{208}_{82}Pb}$ nucleus.
\end{abstract}

\tableofcontents 

\section{Introduction}
\label
{Introduction}
The idea of the atomic nucleus as a collection of protons and neutrons, collectively known as nucleons, which form a bound system due to the exchange of mesons, goes back to Iwanenko \cite{Iwanenko}, Heisenberg \cite{Heisenberg}, Tamm \cite{Tamm}, and Yukawa \cite{Yukawa}.
The simplest renormalizable, charge-independent,  $CP$ and Lorentz invariant Lagrangian governing a system of nucleons and pions reads~\footnote{This Lagrangian defines the standard Yukawa meson theory \cite{Gasiorowicz}.
For simplicity, we have omitted the terms with the vector mesons $\rho$, $\omega$, and $\phi$, responsible for the Hofstadter mutual repulsion of nucleons separated from each other by distances less than 0.3 fm \cite{hofstadter}.} 
\begin{equation}
{\cal L}={\bar\psi}\left(i\gamma^\alpha\partial_\alpha-M\right){\psi}
+\frac12\left(\partial_\alpha\boldsymbol{\phi}\cdot\partial^\alpha\boldsymbol{\phi}-
m_\pi^2\boldsymbol{\phi}\cdot\boldsymbol{\phi}\right)
+g\,{\bar\psi}\gamma_5\boldsymbol{\tau}{\psi}\cdot\boldsymbol{\phi}
+\frac{\lambda}{4}\left(\boldsymbol{\phi}\cdot\boldsymbol{\phi}\right)^2,
\label
{Yukawa-Lagrangian}
\end{equation} 
where $M$ and $m_\pi$ denote, respectively, the nucleon and pion masses (the difference in the masses of particles belonging to the same isotopic multiplet is taken to be negligible), $\boldsymbol{\tau}=\left(\tau_1,\tau_2,\tau_3\right)$ is an isovector whose components are Pauli matrices.
When the Yukawa term responsible for the interaction between nucleons and pions is written explicitly, it takes the form:
\begin{equation}
g\left[\left({\bar\psi}_p\gamma_5{\psi}_p-{\bar\psi}_n\gamma_5{\psi}_n\right){\pi}^0
+\sqrt2\left({\bar\psi}_p\gamma_5{\psi}_n\,{\pi}^-+{\bar\psi}_n\gamma_5{\psi}_p\,{\pi}^+\right)\right].
\label
{Yukawa-interaction}
\end{equation} 
Here ${\pi}^\pm=\frac{1}{\sqrt2}\left({\phi}_1\pm i{\phi}_2\right)$ are the charged pion fields, and ${\pi}^0={\phi}_3$ is the neutral pion field.

A ``technical'' difficulty of this theory is that the estimated couplings $g^2/4\pi\simeq 15$ and $\lambda^2\simeq 1$  \cite{Gasiorowicz} are unsuitable as small parameters for perturbative calculations.
To overcome this difficulty, highly nonlinear Lagrangians for pions and low-energy nucleons consistent with spontaneously broken chiral invariance were invented, and the expansion in coupling constants was replaced with expansions in powers of momentum transfer \cite{Weinberg1990}~\footnote{For a review see \cite{Epelbaum} and \cite{Machleidt}.}.
Unlike the standard Yukawa meson theory (\ref{Yukawa-Lagrangian}), this  approach defies mathematical formulation via a single formula; instead, there is an algorithm for the sequential determination of the suitable terms of chiral perturbation theory stipulated by some additional prescriptions.
To illustrate, we turn to the leading-order Lagrangian of chiral perturbation theory in the so-called heavy baryon formalism \cite{Machleidt}, which allows to get rid of the kinematical dependence on the nucleon mass $M$, 
\begin{eqnarray}
{\cal L}=\frac12\left(\partial_\mu\boldsymbol{\phi}\cdot\partial^\mu\boldsymbol{\phi}
-m_\pi^2\boldsymbol{\phi}^2\right)
\nonumber\\
+\frac{1}{2f_\pi^2}\left[(1-4\alpha)(\boldsymbol{\phi}\cdot\partial^\mu\boldsymbol{\phi})(\boldsymbol{\phi}\cdot\partial_\mu\boldsymbol{\phi})
-2\alpha\boldsymbol{\phi}^2\partial^\mu\boldsymbol{\phi}\cdot\partial_\mu\boldsymbol{\phi} 
+\frac{8\alpha-1}{4}\,m_\pi^2\boldsymbol{\phi}^4\right] 
\nonumber\\
+{\bar N}\left\{i\partial_0 
-\frac{1}{2f_\pi}\boldsymbol{\tau}\cdot\left[g_A({\vec\sigma}\cdot{\vec\nabla}\boldsymbol{\phi}) +\frac{1}{2f_\pi}(\boldsymbol{\phi}\times \partial_0\boldsymbol{\phi})
\right]\right\}N
\nonumber\\
+\frac{g_A}{2f_\pi^3}{\bar N}\left\{\frac12(4\alpha-1)(\boldsymbol{\tau}\cdot\boldsymbol{\phi})\left[\boldsymbol{\phi}\cdot({\vec\sigma}\cdot
{\vec\nabla})\boldsymbol{\phi}\right]
+\alpha\boldsymbol{\phi}^2\left[\boldsymbol{\tau}\cdot({\vec\sigma}\cdot{\vec\nabla})
\boldsymbol{\phi} \right]
\right\}N
\nonumber\\
-\frac12\left[C_S{\bar N}N{\bar N}{\bar N}+C_T({\bar N}{\vec\sigma}N)\cdot({\bar N}{\vec\sigma}{\bar N})
\right]+\ldots
\label
{chiral-Lagrangian}
\end{eqnarray}
Here, $f_\pi$ is the pion decay constant, $N$ represents the large/upper component of the Dirac wave function $\psi$, $\alpha$ is an arbitrary parameter (the $\alpha$ dependence of diagrams with three or four pions entering in vertices of this diagram is expected to drop out), $\vec\sigma$ is a vector whose components are Pauli matrices acting on the Dirac components of the heavy nucleon $N$, $C_S$ and $C_T$ are unknown constants which are determined by a fit to the $NN$ data, and the ellipsis stands for terms that are irrelevant for the derivation of nuclear forces up to first order.
It follows that even Yukawa's fundamental premise that interactions between nucleons occur only through meson exchanges is sacrificed to computability and the rule that two-nucleon forces  $2NF$ dominate over multi-nucleon forces: $2NF\gg 3NF\gg ...$
Indeed, the four-fermion contact terms $\left({\bar N}N\right)^2$ and $\left({\bar N}\vec{\sigma}N\right)^2$, detached from pions, are obligatory for this approach (they renormalize loop integrals).

This approach is successful in correlating many aspects of nuclear phenomenology, but it leaves a feeling of aesthetic dissatisfaction when separate treatment of infrared and ultraviolet effects in quantum chromodynamics (QCD) is compared with the situation in the electroweak interaction theory, for which the conventional expansions in coupling parameters are allowable in the entire energy range up to the grand unification point.

The next difficulty is a disparity between the bond of identical nucleons and that of nonidentical nucleons.
Indeed, Eq.~(\ref{Yukawa-interaction}) shows that the former is due to the exchange of neutral pions $\pi^0$, and the latter is realized through the mediation of charged pions $\pi^+$ and  $\pi^-$.
Is it true that only charged pions are exchanged between a proton and a neutron, turning a proton into a neutron and a neutron into a proton? 
Are interactions between protons and neutrons via $\pi^0$ really forbidden?
There is no fundamental reason why a neutral pion should not be involved in proton-neutron interactions.
This feature of the conventional Yukawa Lagrangian (\ref{Yukawa-Lagrangian}) is absent from its chiral invariant extension (\ref{chiral-Lagrangian}).
Note, however, that empirical evidence of $\pi^0$ meson exchanges between protons and  neutrons within nuclei remains controversial.

The static solution of the Yukawa theory, the famous Yukawa potential 
\begin{equation}
{\phi}(r)=-\frac{g}{4\pi}\,\frac{\exp\left({-m_\pi r}\right)}{r}\,,
\label
{Yukawa-potential}
\end{equation}
makes it clear that the distance at which the Yukawa force is still felt is limited by the Compton wavelength of the pion, $\lambdabar_\pi\approx 1.5$ fm.
Such short-range forces are capable of binding only nearby nucleons.
The well-established phenomenological relationship \cite{Angeli}
\begin{equation}
{R}={R_0}\, A^{1/3}\,,
\label
{R-nucleus}
\end{equation}  
where $R$ and ${A}$ are, respectively, the size and mass number of a given nucleus, and ${R_0}$ is an adjustable parameter, ${R_0}\approx 1$ fm, shows that the greater ${A}$, the larger the nucleus. 
Thus, all heavy nuclei must decay due to the long-range repulsion of the interproton Coulomb forces, unless there is a long-range attraction term of nuclear forces.
However, neither the Lagrangian (\ref{Yukawa-Lagrangian}) nor its chiral invariant extension
(\ref{chiral-Lagrangian}) give rise long-range attraction forces.

A strange thing is that perfect nuclear stability ends at the ${\rm {}^{208}_{82}Pb}$ nucleus,  which has a proton to neutron ratio $Z/N$ of $Z/N\approx 0.65$, rather than at a nucleus for which Coulomb repulsion is more effective, say, at the ${}^{40}_{20}{\rm Ca}$ nucleus, which has $Z/N=1$.

\begin{figure}[htb]
\centerline{\includegraphics[width=0.7\textwidth]{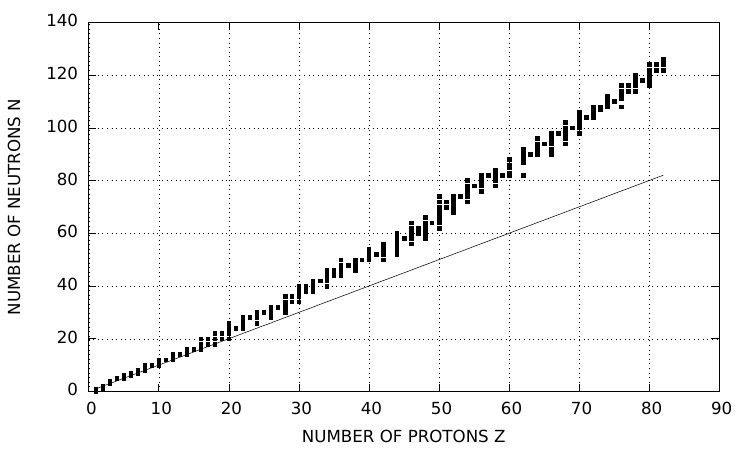}}
\caption{Stable nuclides composed of $Z$ protons and  $N$ neutrons}
\label{stable}
\end{figure}

Every light element of the periodic table has a stable isotope with equal numbers of protons and neutrons~\footnote{The exceptions are beryllium, fluorine, sodium, aluminum, and phosphorus, which have only one stable isotope, namely ${\rm {}^{9}_4 Be}$ with $Z/N=0.8$, ${\rm {}_{9}^{19} F}$ with $Z/N=0.9$, ${\rm {}^{23}_{11} Na}$ with $Z/N\approx 0.92$, ${\rm {}^{27}_{13} Al}$ with $Z/N\approx 0.93$, ${\rm {}^{31}_{15} P}$ with $Z/N\approx 0.94$, respectively.}, whereas stable nuclei heavier than ${\rm {}^{40}_{20}Ca}$ are neutron-rich, and  $Z/N$ diminishes as $A$ grows (c.f. Fig.~\ref{stable}). 
What is the reason for this nuclear composition? 
Why do nuclei consisting only of neutrons not form under terrestrial conditions?~\footnote{Note that neutron stars can be viewed as giant nuclei containing only neutrons. 
However, our concern here is with nuclear physics in a region free from the interference of strong gravity.}
Yukawa's paradigm, embodied in Eq.~(\ref{Yukawa-Lagrangian}) does not provide answers to these questions, and the same is true of Eq.~(\ref{chiral-Lagrangian}) and next terms of  chiral perturbation theory.

The nuclear shell model \cite{Maria}, designed to elucidate the nuclear energy spectrum, was patterned after the conventional quantum-mechanical procedure aimed at elucidating atomic spectra.
With Yukawa's idea that nuclear binding is achieved by meson exchanges, it is natural to expect that the transition of a nucleon from a higher to a lower level is accompanied by the emission of a meson, similar to the accompanying process of photon emission in atomic physics. 
However, no spontaneous emission of pions from nuclei has been observed.
In lieu of the pion emissions, we see $\gamma$-ray emissions.

All these facts indicate that the Yukawa picture of nuclear binding has a very limited range of applicability, the boundaries of which are dictated by QCD.

With the advent of QCD, much effort was directed towards describing nuclei using quarks.
There is direct experimental evidence that a quark located in a nucleus does not have to live in a ``triple room''.
In fact, a free neutron, when it combines with a proton to form a deuteron, loses responsibility for the future fate of its quarks. 
This is easily seen from the fact that a mean lifetime of a $d$ quark relative to $\beta$-decay,  ${\rm T}_{1/2}\approx 896$ s, inside a free neutron, increases to ${\rm T}_{1/2}=\infty$ when this quark is confined to a deuteron.

The simplest way to introduce quarks into nuclear physics is to think of a nucleus of mass number ${A}$ as a bound system of ${\cal N}=3{A}$ quarks placed in a bag of size $R$, 
extending the bag model, well suited to describing hadrons \cite{Chodos}--\cite{DeTar}, to the analysis of nuclei~\cite{Petry}.
A nice feature of this approach is that it attempts to elucidate the structure of hadrons and nuclei on an equal footing by considering bound systems composed of ${\cal N}=3,6,9,...$ quarks.
For such systems it is quite natural that the transition of a quark from an upper energy level to a lower one is accompanied by the emission of a $\gamma$-quantum.
Contrary to the popular belief that the effective degrees of freedom in low-energy QCD are pions (Goldstone bosons of broken chiral symmetry) and nucleons, pions seem to be completely alien to nuclear physics as fundamental building blocks.
Spontaneous emission of pions by nuclei is actually absent, unless we take into account the decay products of the quark-gluon plasma that arises from the head-on collision of heavy relativistic  nuclei.

Meanwhile, for the bag to be stable, ${\cal N}$ and ${A}$ must be related by $R\propto{\cal N}^{1/4}$ \cite{Close}, contrary to Eq.~(\ref{R-nucleus}), and  this discrepancy is especially significant for heavy nuclei.
In addition, the magnetic moments of nuclei differ from those of the bags describing these nuclei \cite{Arima}, \cite{Talmi}.

Attempts to describe nuclei by eliminating gluon degrees of freedom have had some success~\cite{Maltman}, but have been limited to  the lightest nuclei.

To avoid the difficulties of the Yukawa paradigm it is advisable to use an effective theory to low-energy QCD, adapted to research in nuclear physics.
One day we will be able to derive it from first principles, but in the meantime, we must deal with quark-based models that share a number of traits with the desired effective theory.  
Our concern here is with solvable models.
It transpires in Sec.~\ref{Fermi gas model}  that the Fermi gas model is a simple and reliable guide to the composition of stable light nuclei.
Section~\ref{systems of quarks} develops a quark field model amenable to analytical and numerical studies \cite{KPV-1}, \cite{KPV-2}. 
To enhance the predictive power of this model, a refined version of gauge/gravity duality has been proposed in \cite{KPV-3}.
We explicate the basics of this proposal in Sec.~\ref{general holographic}. 
This tool is used in Sec.~\ref{glueball} to predict the main decay channels of the lightest glueball \cite{KPV-4}.
Duality provides an insight into the existence of a maximum allowable electric charge $Z_{\rm max}$ in stable heavy nuclei, and moreover, makes it possible to establish that $Z_{\rm max}\approx 82$ \cite{KPV-5}.
Section~\ref{Holographic-} sketches out this result.
Section~\ref{Conclusion} summarizes the present consideration.

This paper refines, clarifies and in part corrects the ideas and  findings of Refs.~\cite{KPV-1}--\cite{KPV-5}.

\section{The Fermi gas model}
\label
{Fermi gas model}
The Fermi gas model provides a clear explanation of the composition of stable light nuclei.
The decisive argument in favor of using the Fermi gas model is found in  the semiempirical Weizs\"acker formula, which expresses the binding energy per nucleon $B/A$ through the mass number $A$,  number of protons $Z$, and number of neutrons $N$ in a given nucleus,
\begin{equation}
\frac{B}{A}=\alpha-\beta\,A^{-1/3}-\gamma\, \frac{(Z-N)^2}{A^2}-\delta\,Z^2 A^{-4/3}\,.
\label
{weizsacker}
\end{equation}               
The coefficient  $\gamma$ of the term associated with the effect of mutual repulsion of identical fermions due to the Pauli exclusion principle is 32 times greater than the coefficient $\delta$ of the term responsible for the effect of Coulomb interproton repulsion. 
Therefore, the degeneracy pressure plays a crucial role in the balance of forces, while the Coulomb interaction is of much lesser importance and can be neglected as a first approximation. 

We are thus able to develop an ab initio model of the Fermi gas by imagining that the nucleus is a mixture of two Fermi gases, one containing $u$ quarks and the other $d$ quarks. 
The degeneracy pressure $P$ in this system varies directly as
\begin{equation}
n_u^{5/3}m_u^{-1}+
\left({\cal N}-n_u\right)^{5/3}m_d^{-1}\,,
\label
{deg_pressure}
\end{equation}              
where $n_u$  and $n_d={\cal N}-n_u$  are the numbers of  $u$ and $d$ quarks, respectively,  $m_u$ and $m_d$ are their masses, and ${\cal N}$ is the total number of quarks in the nucleus.
It is then clear why stable light nuclei are half-filled with $d$ quarks~\footnote{More exactly, the set of stable isotopes of every light element of the periodic table contains a stable nucleus with $n_u=n_d$, except for five light elements, each of which is represented by a single stable nuclide, namely ${\rm {}^{9}_{4}Be}$, with $n_d/n_u\approx 1.077$, ${\rm {}_{9}^{19} F}$ with $n_d/n_u\approx 1.036$, ${\rm {}^{23}_{11} Na}$ with  $n_d/n_u\approx 1.029$, ${\rm {}^{27}_{13} Al}$ with  $n_d/n_u\approx 1.025$, and ${\rm {}^{31}_{15} P}$ with  $n_d/n_u\approx 1.022$.}.
Since $P$ is proportional to the energy density ${\cal E}$, which must be minimal in order for the system to be  stable, we find 
\begin{equation}
n_u={\cal N}\left(m^{3/2}_u+m^{3/2}_d\right)^{-1} m^{3/2}_u\,.
\label
{n-u-m-d-n-tot-over-m-tot}
\end{equation} 
If we assume that $m_u\approx m_d$, which is the case for the constituent quark masses, then Eq.~(\ref{n-u-m-d-n-tot-over-m-tot}) becomes  $n_u\approx \frac12\,{\cal N}$, that is,                           
\begin{equation}
n_u\approx n_d\,.
\label
{n-approx-n}
\end{equation}   
This rule explains the composition of stable light nuclei.  
For example,  $n_u=n_d=3$ in a deuteron ensures its stability, while $n_u=4$ and $n_d=5$ in a triton is consistent with its instability.
Stability is even more absent in the dineutron, which contains $n_u=2$ and $n_d=4$.
The quark composition of ${\rm {}^{9}_{4}Be}$ can be considered in reasonable agreement  with Eq.~(\ref{n-approx-n}), if we take into account that $n_u=13$ is not so far away from $n_d=14$: $n_d/n_u\approx 1.077$, and this is much more true in relation to  ${\rm {}_{9}^{19} F}$ for which  $n_u=28$ and $n_d=29$: $n_d/n_u\approx 1.036$, ${\rm {}^{23}_{11} Na}$  for which  $n_u=34$ and $n_d=35$:  $n_d/n_u\approx 1.029$, ${\rm {}^{27}_{13} Al}$  for which  $n_u=40$ and $n_d=41$:  $n_d/n_u\approx 1.025$, and ${\rm {}^{31}_{15} P}$  for which  $n_u=46$ and $n_d=47$:  $n_d/n_u\approx 1.022$.

This argument, repeated in terms of protons and neutrons, gives $Z\approx N$.
However, the statistical physics methods underlying the Fermi gas model are poorly suited to the description of few-particle systems, so that $N/Z$ for some stable light isotopes turns out to be quite rough, for example,  $N/Z= 1.25$ for ${\rm {}^{9}_{4}Be}$.
Moreover, the very rationale behind this model fails for the deuteron viewed as a system of two non-identical nucleons, because there is nothing in this system to be affected by the degeneracy pressure.

It transpires why it is impossible to assemble neutrons into stable bound systems: in such systems, $n_d=2n_u$, which is the utmost deviation of ${\cal E}$  from the minimum.

For  $Z>20$, the rule (\ref{n-approx-n}) no longer holds.
The ratio $n_d/n_u$ increases approximately linearly with increasing $Z$, as seen in  Fig.~\ref{stable}. 
What is the reason for abandoning the requirement that ${\cal E}$ should be minimal for stable nuclei heavier than  ${\rm {}^{40}_{20}Ca}$? 
The qualitative explanation is this.
The heavier the nucleus, the more widely separated quarks it contains. 
When two quarks are far apart, infrared QCD effects come to play, dramatically enhancing their mutual attractions. 
To balance the forces, it is necessary to increase the degeneracy pressure, which implies the necessity of deviating from the minimum energy density ${\cal E}$. 
A quantitative analysis of this phenomenon will be given below, within the framework of a modified bag model.

\section{Towards the effective theory to low-energy QCD}
\label
{systems of quarks}
20 years ago the bag model was a very respectable research topic ({the interested reader may consult Ref.~\cite{DeTar} where the MIT bag model \cite{Chodos}, SLAC bag model \cite{Bardeen}, soliton bag model \cite{Lee}, and their developments are considered at length}).
Another phenomenological model of quarks confined to hadrons with the aid of the Cornell potential \cite{Cornell75},
\begin{equation}
{V_{\rm C}(r)}=-\frac{\alpha_s}{r}+\sigma r \,,
\label
{Cornell}
\end{equation}
has also enjoyed popularity for years \cite{GodfreyIsgur}--\cite{Barnes}.
The motivation for Eq.~(\ref{Cornell}) is as follows.
The first term grasps the interaction of quarks when they come close to each other.  
To find it, we need to take into account the asymptotic freedom at short distances, which allows us to use perturbation theory and calculate the one-gluon exchange term.
Then ${\alpha_s}$ is regarded as a coupling parameter of the strong interaction between quarks~\footnote{Recent theoretical and experimental progress in understanding the $\alpha_s$ coupling can be found in the semi-popular review \cite{Brodsky} and the references to original works therein.}.
The linearly rising term is attributed to the hypothetical mechanism of squeezing the gluon field lines into a thin tube with constant energy per unit length.
The relevance of this term is usually justified by lattice QCD calculations of Wilson loops \cite{Wilson}--\cite{Bander}.

Both models discussed were combined in \cite{KPV-1}, \cite{KPV-2}, \cite{KPV-5} in the hope of obtaining a {modified} bag model suitable for describing nuclei. 

\subsection{Modified bag model}
\label
{modified bag model}
A central idea of the modified bag model is that a single quark in a given nucleus, driven by a mean field generated by all other constituents of the nucleus, is responsible for the static properties of the nucleus (a similar idea underlies the nuclear shell model \cite{Maria} 
with the obvious replacement of quarks by nucleons).
The quark is described by a spinor field  $\Psi$ whose dynamics is assumed to be governed by
the action
\begin{equation}
{\cal S}=\int d^4x\,{\overline\Psi}\left[\gamma^\alpha\left(i\partial_\alpha+g_VA_\alpha\right) +g_S\Phi-m_0\right]\Psi \,,
\label
{QCD-Lagrangian}
\end{equation}                        
where  $A_\alpha=(A_0,-{\bf A})$ and $\Phi$ are, respectively, the Lorentz vector and scalar potentials of the mean field, $g_V$ and $g_S$ are their associated coupling constants, and $m_0$ is the current quark mass. 
What is the motivation for taking this action?
It is well known \cite{Wheeler} that  if the carrier of interaction is a scalar field or a symmetric  tensor field of the second rank, then particles with charges of the same sign are attracted.
An example is gravity.
However, if the mediator is a vector field, then particles with the same charge sign experience mutual repulsion \cite{Wheeler}.
This is the case with electromagnetism.
The presence of vector and scalar potentials, $A_\alpha$ and $\Phi$, in the action (\ref{QCD-Lagrangian}) is necessary to balance the interquark forces.
Recall that the $u$ quark has a positive electric charge, $q=\frac23 e$, the $d$ quark has a negative electric charge, $q=-\frac13 e$, and each quark is endowed with the same baryon charge $B=\frac13$.
The potentials $A_\alpha$ and $\Phi$ can, in principle, be found by averaging the electromagnetic and gluon fields over all their sources in the nucleus.
However, this is an extremely difficult task.
We therefore relegate $A_\alpha$ and $\Phi$ to the status of the mere background fields with a fixed space dependence like that shown in (\ref{Cornell}).

The study of the modified bag model begins with the simplest case of non-relativistic motion of a quark in a spherically symmetric mean field, so that in the vector potential $A_\alpha(x)$ it is sufficient to use only its time component $A_0(r)$. 
This is consistent with the fact that the ground state nucleus is roughly spherical, and that the Pauli exclusion principle, acting through the already filled orbitals, suppresses the role of long-range correlations  \cite{Walecka}.
Thus, the Dirac Hamiltonian
\begin{equation}
H_D=\boldsymbol{\alpha}\cdot{\bf p}+{\mathbb I}\,U_V({r})+\beta\left[m_0+U_S({r})\right],
\label
{Dirac-Hamiltonian}
\end{equation}
where $\boldsymbol{\alpha}$ and $\beta$ are the Dirac matrices, ${\mathbb I}$ is an identity matrix, ${\bf p}=-i\nabla$ is the momentum of the quark, and $U_V=g_VA_0$, $U_S=g_S\Phi$, is taken as the starting point~\footnote{The use of a Hamiltonian of this type is common practice in the nuclear shell model; and it was previously proposed in \cite{Page} to describe the quark in the nucleus.}. 

The question may arise of whether it is not superfluous to use the Dirac equation to describe a nonrelativistic quark.
In response, we note that this equation is not necessarily associated with relativistic dynamics.
What is more important for us is that this equation constitutes an exact link between the quark spin degrees of freedom and the mean field, which frees us from worrying about the correct form of the spin-orbit coupling term.
In addition, this equation is a flexible tool, best suited to account for exact and broken symmetries inherent in nuclear physics.

Searching for solutions of the Dirac equation leads to the eigenvalue problem 
\begin{equation}
H_D\Psi=\varepsilon\Psi\,.
\label
{Dirac-genera}
\end{equation}
To complete the definition of this problem the following two conditions should be added:

\noindent
(i) Pseudospin symmetry condition \cite{Ginocchio97}~\footnote{Alternatively, $U_{V}$ and $U_{S}$ are subject to the spin symmetry condition
\noindent
\begin{equation}
U_S(r)=U_V(r)+{\tilde{\cal C}}\,.
\label
{spin}
\end{equation}
We will briefly touch on this alternative, but our main focus will be on the pseudospin symmetry condition.
For an extended discussion of the pseudospin and spin symmetry conditions see Refs.~\cite{Ginocchio} and \cite{Liang}.}:
\begin{equation}
U_S(r)=-U_V(r)+{\cal C}\,.
\label
{pseudospin}
\end{equation}
Here, ${\cal C}$ is a positive constant defined for each type of nuclei, 

\noindent
(ii) Asymptotic condition: $|U_{V}(r)|$,  $|U_{S}(r)|$  grow in space.

Substituting (\ref{pseudospin}) into (\ref{Dirac-Hamiltonian}) gives
\begin{equation}
H_D=\boldsymbol{\alpha}\cdot{\bf p}+U_V(r)({\mathbb I}-\beta)+\beta\left(m_0+{\cal C}\right).
\label
{Dirac-Hamiltonian-spin}
\end{equation}
This is a transformed Dirac Hamiltonian with a shifted mass, 
\begin{equation}
m_0\to m=m_0+{\cal C}\,.
\label
{mass_shift}
\end{equation}
It is as if the pseudospin symmetry condition  (\ref{pseudospin}) converts the bare quark mass  $m_0$ into the dressed quark mass $m$, while the pseudospin symmetry condition becomes  $U_S=-U_V$.

\begin{figure}[htb]
\centerline{\includegraphics[width=0.6\textwidth]{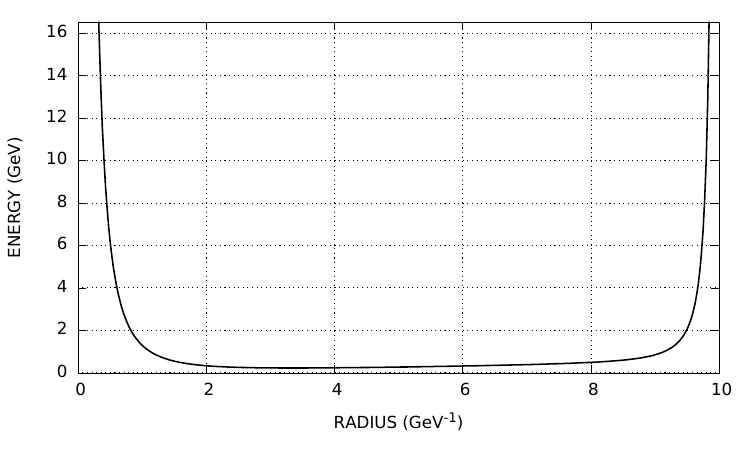}}
\caption{Effective potential  ${U}\left(r;\varepsilon\right)$ }
\label{pseudo-spin-potential}
\end{figure}

To solve Eq.~(\ref{Dirac-genera}), we separate variables in the usual way \cite{Ginocchio}. 
The radial part is
\begin{equation}
{f'}+\frac{1+\kappa}{r}\,f-\xi g=0\,,
\label
{Dirac_radia_f}
\end{equation}
\begin{equation}
{g'}+\frac{1-\kappa}{r}\,g+\eta f=0\,.
\label
{Dirac_radia}
\end{equation}
Here, $f$ and $g$ denote,  respectively, the upper and lower radial components of the Dirac bispinor, $\kappa=\pm(j+\frac12)$ are eigenvalues of the operator ${K}=-\beta\,({\boldsymbol{\tau}}\cdot{\bf L}+1)$ (where $\boldsymbol{\tau}$ are the Pauli matrices and ${\bf L}=-i{\bf r}\times\nabla$), which 
commutes with the spherically symmetric Dirac Hamiltonian  (\ref{Dirac-Hamiltonian}), and
\begin{equation}
\xi(r)=\varepsilon+{m}+U_S(r)-U_V(r)\,,
\label
{a-df}
\end{equation}
\begin{equation}
\eta(r)=\varepsilon-{m}-U_S(r)-U_V(r)\,.
\label
{b-df}
\end{equation}
Expressing $g$ through $f$ and  $f'$, we arrive at a one-dimensional Schr\"odinger-like
equation
\begin{equation}
F''+k^2F=0\,,
\label
{1D_Schroedinger}
\end{equation}                                          
\begin{equation}
k^2={\varepsilon^2-m^2}-U(r;\varepsilon)\,.
\label
{k-df}
\end{equation}
The quantity $U(r;\varepsilon)$ is called the {\it effective potential}.
Taking $U_V=\frac12 V_{\rm C}$ and substituting  $U_S=-U_V$ into (\ref{a-df}) and (\ref{b-df}), we find 
\[U(r;\varepsilon)=\frac{\kappa(\kappa+1)}{{r^2}}
+\left(\varepsilon-{m}\right)V_{\rm C}(r)
\]
\begin{equation}
+\frac{1}{r^2}\left\{\frac{3\left(\alpha_s+\sigma r^2\right)^2}
{4\left[\sigma r^2-\left(\varepsilon +{m}\right)r-\alpha_s\right]^2}+
\frac{\alpha_s\left(\kappa+1\right)+\kappa\sigma r^2}
{\sigma r^2-\left(\varepsilon+{m}\right)r-\alpha_s}
\right\}.
\label
{U_eff-PSEUDO}
\end{equation}
The last two terms in the expression (\ref{U_eff-PSEUDO}) are singular at 
\begin{equation}
r_{\ast}=\frac{\left(\varepsilon +{m}\right)+
\sqrt{\left(\varepsilon +{m}\right)^2+4\sigma \alpha_s}}{2\sigma}\,.
\label
{sol}
\end{equation}
The form of ${ U}\left(r;\varepsilon\right)$ with some values of $m$, $\varepsilon$, ${\alpha_s}$, and $\sigma$  is shown in Fig.~\ref{pseudo-spin-potential}. 

The most important property of the effective potential  ${U}\left(r;\varepsilon\right)$  is that it develops a singularity,
\begin{equation}
{U}\left(r;\varepsilon\right)\sim \gamma\left(r-r_\ast\right)^{-2},
\quad
\gamma>0\,,
\label
{effective_potential}
\end{equation}
at a finite radius $r_\ast$.
We thus see that the pseudospin symmetry condition (i), together with the asymptotic condition (ii), greatly enhances the interaction between the spin degrees of freedom of the quark and the mean field, resulting in a spherical cavity whose surface has an infinite effective potential, ${U}\left(r_\ast;\varepsilon\right)=\infty$.
It seems natural to identify the cavity with the interior of the described nucleus, and to equate the radius of the cavity $r_\ast$ with the radius of the nucleus $R$.

It is straightforward to show \cite{KPV-1} that the quark ends up in an infinitely deep potential well of almost rectangular shape, similar to that shown in Fig.~\ref{pseudo-spin-potential}, whenever  $|U_{V}(r)|$ and $|U_{S}(r)|$  increase monotonically with $r$, and the pseudospin symmetry condition is satisfied.
We are thus free to vary the forms of $U_{V}(r)$ and $U_{S}(r)$ over a wide range to find the
best match with the observed properties of the nucleus.

A procedure similar to the one described above, replacing the upper component of the Dirac bispinor with the lower one and the pseudospin symmetry condition with the spin symmetry condition, leads to essentially the same results.

The criterion for {quark confinement} can now be extended from hadrons to larger systems containing $3A$ quarks as follows: the probability amplitude that the quark is somewhere inside the cavity of radius $r_\ast$ is given by the solution of the Schr\"odinger-like equation (\ref{1D_Schroedinger}), and  outside this cavity it is equal to 0. 
A justification for this criterion follows from the theorem according to which the tunneling of a particle through an  infinitely high one-dimensional barrier of the form $\gamma\left(r-r_\ast\right)^{-2}$ is forbidden if and only if $\gamma\ge \frac34$ \cite{Dittrich}.
The barrier produced by the effective potential ${U}\left(r;\varepsilon\right)$ defined by Eq.~(\ref{U_eff-PSEUDO}) satisfies this condition.

An attractive feature of the modified bag model is that it dispenses with the artificial elements of the standard bag model (such as surface tension, vacuum pressure, and the Heaviside step function $\theta_V$) that were introduced to stabilize the bag geometry \cite{DeTar}.

\subsection{The static properties of nuclei }
\label
{bound_systems_of_quarks}
The modified bag model provides a unified description of the  sequence of stable nuclides in the periodic table.
Their properties are linked to those  of a discrete series of stationary solutions of the Schr\"odinger-like equation (\ref{1D_Schroedinger}).
Each eigenvalue $\varepsilon_{n}$ of the Hamiltonian (\ref{Dirac-Hamiltonian-spin}) is  identified with the mass of a stable nucleus whose position in this sequence is determined by the radial and spin-orbit quantum numbers,  ${n}$ and $\kappa$.
The size of this nuclide $r_\ast$  is determined by $\varepsilon_{n}$ via Eq.~(\ref{sol}).
For the numerical solution of Eq.~(\ref{1D_Schroedinger}), we used the following parameters:
$\alpha_s=0.7$, $\sigma=0.1\,{\rm GeV}^2$ (as used in description of quarkonia), and $m=0.33$ GeV.
It has been shown that the eigenvalues $\varepsilon_{n}$ are proportional to $\sqrt{n}$; assuming that ${n}$ equals the integer part of ${A}^{2/3}$,  the cavity radii $r_\ast$  then scale as $R_0{A}^{1/3}$, with $R_0\approx 1$ fm, in good agreement with Eq.~(\ref{R-nucleus}).
To test whether the predicted ${n}=\lbrack{A}^{2/3}\rbrack$ relation reproduces the variation of other properties  along the sequence of stable nuclides, we compared   the magnetic dipole moment of a quark driven by the mean field of a given nuclide with the experimentally measured magnetic moment of the nuclide itself. 
The calculated results agree with most experimental data within $\sim 10\%$ (with a few $\sim 20\%$ outliers) for a large set of stable isotopes \cite{KPV-2}.

\subsection{Stable nuclei heavier than ${\rm {}^{40}_{20}Ca}$}
\label
{Z>20}
\begin{figure}[htb]
\includegraphics[width=0.49\textwidth]{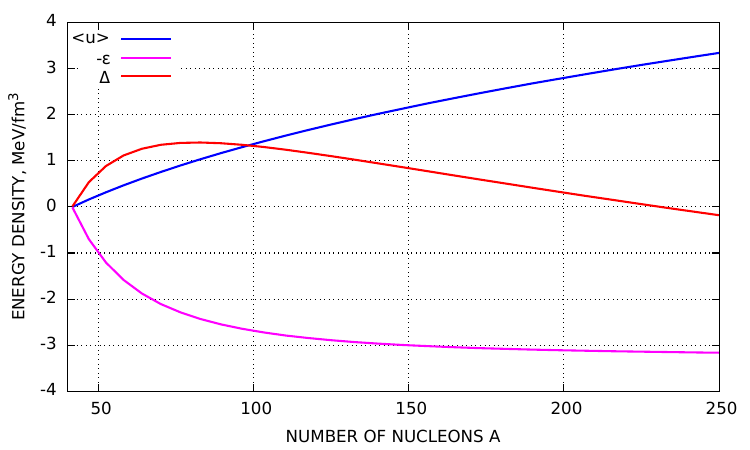}\quad
\includegraphics[width=0.49\textwidth]{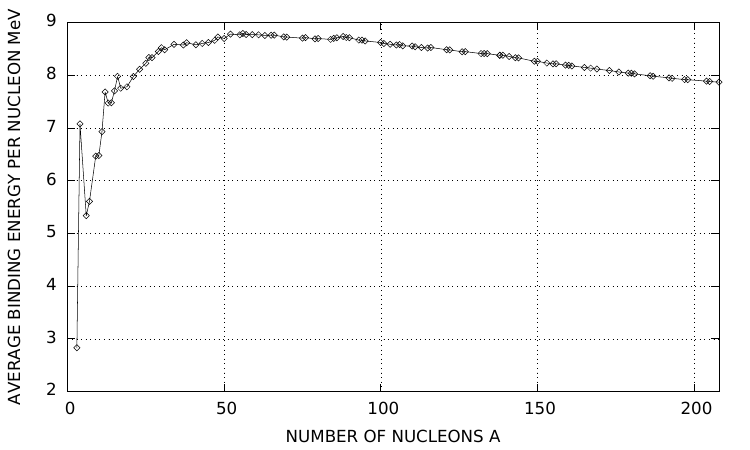}
\caption{Average  mean-field energy density  $\langle{\mathfrak{u}}\rangle$, the 
energy density ${\cal E}$ associated with  degeneracy pressure, 
and their difference $\Delta=\langle{\mathfrak{u}}\rangle-{\cal E}$ (left panel); and the
average binding energy per nucleon, $B/A$ (right panel)}
\label{comparison}
\end{figure}

Let us turn to the balance of forces in stable nuclei from
${}^{40}_{20}{\rm Ca}$ to ${}^{208}_{82}{\rm Pb}$, where the mass number $A$ runs from $40$ to $208$,  corresponding to $12\le n\le 35$.
We take the set of  solutions $F_{n}(r)$ of the Schr\"odinger-like equation (\ref{1D_Schroedinger}), calculate the average energy density $\langle{\mathfrak{u}}\rangle$ of the mean field for each of these states, 
\begin{equation}
\langle{\mathfrak{u}}\rangle
=\frac{3}{4\pi\left(R_0\sqrt{n}\right)^3}\int_0^{r_{\ast}}dr\,{ U}\left(r;\varepsilon_{n}\right)
|F_{n}(r)|^2\,,
\label
{energy density}
\end{equation}
and compare $\langle{\mathfrak{u}}\rangle$ with the energy density ${\cal E}$ associated with  degeneracy pressure \cite{KPV-5}.
We use the empirical fact, evident from Fig.~\ref{stable}, that for nuclei with $Z$ in the range $20\le Z\le 82$, the neutron  excess ${\Delta N}$ varies approximately linearly with $Z$, 
\begin{equation}
\Delta N=0.71\left(Z-20\right)=0.71\left(0.5 A -\Delta N -20\right). 
\label
{Delta N}
\end{equation}
Combining (\ref{Delta N}) with the expression for ${\cal E}$ which is proportional to  ${{Z}^{5/3}}{{M_p}^{-1}}+{\left(A-Z\right)^{5/3}}{M_n^{-1}}$, we obtain     
\begin{equation}
{{\cal E}}\propto 
\left[\left(0.29+\frac{8.3}{n^{3/2}}
\right)^{5/3}\frac{1}{M_p}+\left(0.71-\frac{8.3}{n^{3/2}}\right)^{5/3}\frac{1}{M_n}\right].
\label
{deg_pressure_to_compare}
\end{equation}                                           
Here, $M_p$ and $M_n$ denote the proton and neutron  masses.
The  numerically evaluated values of $\langle{\mathfrak{u}}\rangle$ and ${\cal E}$, as well as their difference,  $\Delta=\langle{\mathfrak{u}}\rangle-{\cal E}$, are plotted in Fig.~\ref{comparison} (left).
The coincident values of $\langle{\mathfrak{u}}\rangle$ and ${\cal E}$  at $A=40$ are used as the reference point, and the subsequent behavior of these quantities is shown. 
It is seen from Fig.~\ref{comparison} (left) that the  interquark QCD attraction does not exactly balance the  mutual repulsion of quarks associated with degeneracy pressure.
How is the balance of forces restored? 
Note that the graph of $\Delta(A)$ closely resembles that of the quantity  $B/A$, the experimentally measured average binding energy per nucleon, Fig.~\ref{comparison} (right).

However, if we abandon the view of the nucleus as a bound system of nucleons,  the concept of binding energy per nucleon becomes meaningless.
The quantity  $B/A$ should instead be interpreted as the mass loss of a dressed quark in a given nucleus relative to its mass in a free proton.
Therefore,  the dressing and re-dressing of quarks are responsible for fine-tuning the balance of forces in the nucleus \cite{KPV-5}.
To gain an intuitive understanding of the mechanism of quark mass loss, let us turn to non-Abelian solutions of the classical ${\rm SU}\left(N\right)$ Yang--Mills--Wong theory \cite{k98}--\cite{K19}.
For a self-interacting particle in the non-Abelian phase described by this theory, the square of its effective mass, which equals the square of the four-momentum, is
\begin{equation}
p^2=m^2\left(1+\ell^2 a^2\right).
\label
{p-mu-dress-quark-sqr}
\end{equation}                                      
Here, $m$ is the renormalized mass of the particle, $a^2$ is the four-acceleration squared~\footnote{Recall that $a^2<0$.}, and $\ell$ is a length characteristic of this particle, $\ell={8}/({3\,g_{\rm {YM}}^2\, m})\!\left(1-{1}/{{N}}\right)$, with $g_{\rm {YM}}$  the Yang--Mills coupling constant. 
It follows that the greater the acceleration of a dressed quark, the smaller its effective mass $\sqrt{p^2}$.
If we regard the proton as the ${\rm {}^{1}_{1}H}$ nucleus at the beginning of the periodic table, then we infer that a quark confined to a region of minimal size,  $r_\ast\approx R_0$, exhibits the most quiescent mode of existence.
As $A$ and the associated $r_\ast$ increase, the quark dynamics become increasingly vigorous.
It is evident from Fig.~\ref{comparison} (right) that the changes in the quark dynamical regime are especially large when passing from one stable nuclide to a neighboring one at the very beginning of the periodic table.

These observations lead  us to conclude that the properties of the dressed quark, in particular its mass, depend on its environment \cite{KPV-5}.
A $d$ quark in a deuteron is not identical to a $d$ quark in a lead nucleus~\footnote{This is not so surprising if we take into account that the $d$ quark in the proton is  $\approx 4.6$ times heavier than the $d$ quark in the $\pi^0$ meson.}, and the same applies to the $u$ quark.
However, all quarks of the same flavor within a given nucleus are identical and indistinguishable.

\section{Holographic nuclear physics}
\label
{general holographic}
To enhance the predictive power of the modified bag model, it is useful to employ  {gauge/gravity duality} \cite{Maldacena}, \cite{Witten}, \cite{Klebanov}~\footnote{Other names for this duality 
include {``the correspondence between the theory of quantum gravity in anti-de Sitter space and the conformal field theory in Minkowski space''}, ``AdS/CFT'', and {``the holographic principle''}.
The main AdS/CFT prescriptions, and computational techniques are reviewed in \cite{Ammon} and \cite{Nastase}.}.
Loosely speaking, this framework posits that much of nuclear and subnuclear physics in Minkowski space ${\mathbb R}_{1,3}$ is captured by the physics of black holes (BHs) in five-dimensional anti-de Sitter space ${\rm AdS}_5$ whose boundary is ${\mathbb R}_{1,3}$.

\subsection{Refined version of gauge/gravity duality}
\label
{novel_holographic}
A popular line of research in gauge/gravity duality is to map a  BH in ${\rm AdS}_5$ onto a 
quark-gluon plasma in ${\mathbb R}_{1,3}$ \cite{Herzog}. 
Another approach pursues the idea of mapping a Dp-brane onto a subnuclear system in the confinement phase \cite{Witten-98}, \cite{Sakai}.
In the present context, it is appropriate to adopt a new version of duality \cite{KPV-3}: ``The state of a Dirac particle inside an {\it extremal} BH located in ${\rm AdS}_5$ is mapped onto the corresponding state of a quark inside a {\it stable} nucleus in ${\mathbb R}_{1,3}$''.
What is special about extremal BHs?
They are not subject to Hawking evaporation.
The extremal BH and the stable nucleus share a common feature in that both lack the ability to spontaneously eject their constituents.

There are two reasons why an unstable nucleus is not dual to an ordinary BH that undergoes  Hawking radiation. 
First, quantum-mechanical processes (decay and recombination, emission and absorption) are {reversible}, whereas Hawking radiation is irreversible. 
Second, quantum-mechanical particles of a given species are identical and indistinguishable.
It is reasonable to demand the same from their duals.
Let us check whether two evaporating BHs can be identical.
Consider two Lorentz reference frames, ${\cal O}$ and ${\cal O}'$, moving at a constant speed relative to each other and meeting at some instant.
Let a Schwarzschild BH be placed in ${\cal O}$, and suppose its mass $M_{{\cal O}}$ at this instant  equals the mass $M_{{\cal O}'}$ of another Schwarzschild BH placed in ${\cal O}'$.
Will the equality $M_{{\cal O}}=M_{{\cal O}'}$ hold in the future?
The evaporation rate, measured with respect to the proper time, is the same for both  BHs.
Therefore, at a later time,  simultaneous measurement of the masses of the BHs placed in ${\cal O}$ and ${\cal O}'$ will yield $M_{\cal O}\not= M_{{\cal O}'}$.  
The relativistic effect of time dilation precludes evaporating BHs from being considered identical objects \footnote{It is easy to see here the manifestation of the notorious twin paradox if we endow the twins at the moment of their separation with identical ``hourglasses'' embodied by evaporating BHs.}.

To simplify our discussion as much as possible, we begin with the case of nonrotating black holes.
For such a BH to be extremal~\footnote{The {\it extremality property} of a BH is as follows: An extremal BH has a {\it single} horizon, or, in more technical terms, a BH is extremal if the equation $g_{00}=0$ has a {\it unique positive} root (i. e., two  positive roots that merge).}, it must be charged, which is clear from the gravitational  potential of this configuration:
\begin{equation}
g_{00}=1-\frac{2M}{r}+\frac{Q^2}{r^2}\,,
\label
{g-00-4D}
\end{equation} 
where $M$ and $Q$ are the mass and electric charge of the BH.
Comparison of (\ref{g-00-4D}) with (\ref{Cornell}) suggests that for the holographic mapping of an extremal BH onto a modified bag to be realized, $g_{00}$ must contain a term that increases with $r$.
This is the case in higher-dimensional spacetimes.
To illustrate,  $g_{00}$ for a non-rotating charged BH in ${\rm AdS}_5$ is given by 
\begin{equation}
g_{00}=1-\frac{2M}{r^2}+\frac{Q^2}{r^4}+\frac{r^2}{L^2}\,,
\label
{f(r)}
\end{equation}
where $L$ is the radius of curvature of ${\rm AdS}_5$ (it is convenient to set $L=1$, thereby fixing the length scale and rendering $r$ a dimensionless variable).

Why is ${\rm AdS}_5$ needed?
This is a five-dimensional one-sheeted hyperboloid in the six-dimensional pseudo-Euclidean space ${\mathbb R}_{2,4}$. 
The hyperboloid is invariant under the group of pseudo-orthogonal transformations ${\rm SO}(2,4)$, which is isomorphic to the group of  conformal  transformations  ${\rm C}\left(1,3\right)$ 
acting on ${\mathbb R}_{1,3}$. 
Thus, the ${\rm AdS}_5$ geometry itself induces conformal symmetry in its holographic image.
If we assume that bare quarks are massless, then the holographic image of the ${\rm AdS}_5$ geometry is a conformal field theory governed by the action (\ref{QCD-Lagrangian})  with $m_0=0$.
However, conditions (i) and (ii) break the conformal invariance of the modified bag model.
Indeed, condition (i) introduces a dimensional parameter ${\cal C}$ which trns the massless bare quark into a massive dressed quark, and condition (ii) implies that the mean field is characterized by dimensional quantities such as $\sigma$, the coefficient of the linearly rising term in (\ref{Cornell}).
On the other hand, if ${\rm AdS}_5$ is ``pocked with black holes'', then the ${\rm SO}(2,4)$ isometry of the hyperboloid is broken.
This means that the geometry of such deformed manifolds induces a breaking of the C(1,3) invariance in their holographic images.
Both the ${\rm SO}(2,4)$ isometry of ${\rm AdS}_5$ and its  breaking due to the presence of black holes are purely geometric phenomena.
This suggests that gauge/gravity duality can provide insight into the mechanism of QCD symmetry breaking characteristic of the infrared region, particularly in nuclear physics.

Let us now examine how this version of duality works.

\subsection{Duality clarifies the pseudospin symmetry condition}
\label
{holographic_explanation}
The metric of a non-rotating charged BH in ${\rm AdS}_5$ is
\begin{equation}
ds^2= h_t^2(r^2)\,dt^2-{h_t^{-2}(r^2)}{dr^2}-r^2\left[d\psi^2+\sin^2\psi\left(d\vartheta^2+\sin^2\vartheta\,d\varphi^2\right)\right],
\label
{RN_AdS}
\end{equation}
where  $h_t^2$ denotes the  $g_{00}$ component of the metric tensor, as given in (\ref{f(r)}).
Imposing the  extremality condition yields the unique event horizon at
\begin{equation}
{r}_\star^2=\left(M+\frac92\,Q^2\right)\!\left(1+6M\right)^{-1}\,.
\label
{r-h}
\end{equation} 

We then supplement these data with a static solution to the five-dimensional Maxwell equations for this geometry,
$A_A=(A_0,0,0,0,0)$, where 
\begin{equation}
A_{0}(r)=\frac{Q}{r^2}\,.
\label
{A_and_A_5}
\end{equation}

\begin{figure}[htb]
{\includegraphics[width=0.49\textwidth]{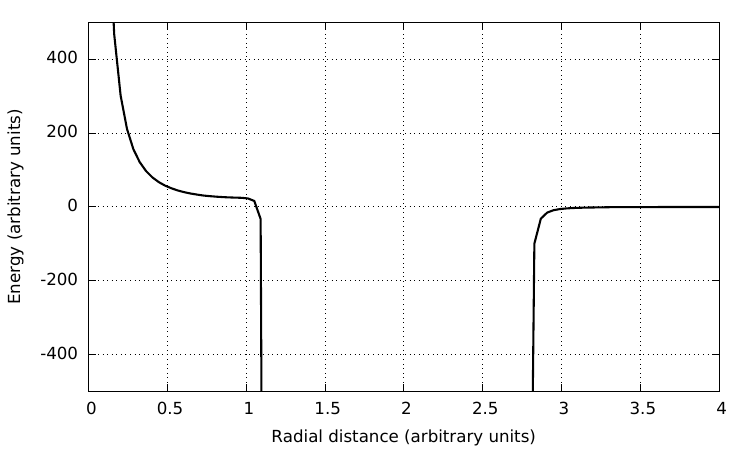}}\quad
{\includegraphics[width=0.49\textwidth]{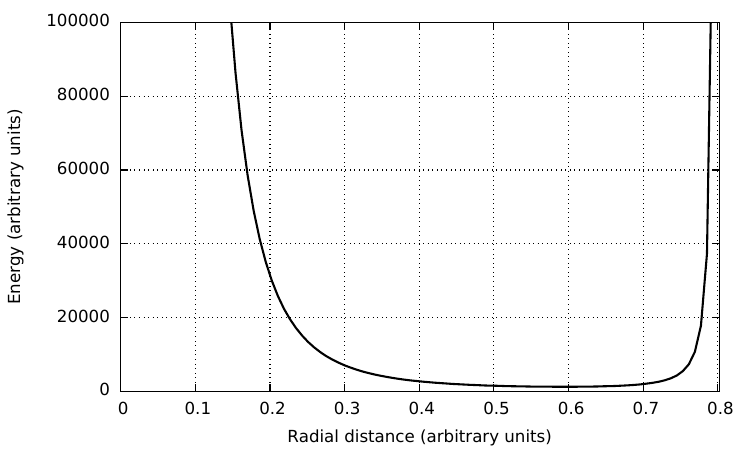}}
\caption{Effective potential ${\cal U}(r;E)$ for a non-extremal  BH (left); and for an extremal BH (right), provided that condition (\ref{equilibr_condition}) is satisfied}
\label{eff-potential-two-horiz}
\end{figure}

The Dirac equation in such gravitational and electromagnetic fields
\begin{equation}
\left[\gamma^A e^\alpha_A\left(\partial_\alpha+\Gamma_\alpha-iqA_\alpha\right)+\mu\right]\chi(x)=0\,
\label
{Dirac_bulk-}
\end{equation}                         
(where $e^\alpha_A$ is a pentad, and $\Gamma_\alpha$ and $A_\alpha$ are gravitational and electromagnetic connections) allows separation of variables \cite{Cotaescu}--\cite{Wu-8}. 
Following the procedure of the previous section, the radial part can be reduced to a 
Schr\"odinger-like equation \cite{KPV-3}.
The resulting effective potential ${\cal U}(r;E)$ is rather complex, and is therefore not presented here.
For a non-extremal BH,  ${\cal U}(r;E)$ would be
highly singular, and the Dirac particle would fall into an infinitely deep potential well, as shown in Fig.~\ref{eff-potential-two-horiz} (left).
\psfrag{x}[c][c][0.5]{${r}$}\psfrag{y}[c][c][0.5]{${\cal U}(r;E)$}
\psfrag{xlabel}[c][c]{\tiny RADIAL DISTANCE (ARBITRARY UNITS)}
\psfrag{ylabel}[c][c]{\tiny ENERGY (ARBITRARY UNITS)}
For an extremal BH, the strongest singularities of ${\cal U}(r;E)$  disappear,  if we impose an additional condition (discussed below), the coefficient of the leading singularity becomes positive, so that ${\cal U}(r;E)$ behaves as shown in Fig.~\ref{eff-potential-two-horiz} (right).

To simplify matters, we consider the radial part of the Dirac equation near the event horizon  (${r}\to{r}_\star$), 
\begin{equation}
\frac{d}{dr^2}
\left(\begin{array}{c} f \\g\end{array}\right)=\frac{\Lambda_0}{{r}_\star^2-r^2}\left(\begin{array}{cc}-\kappa & \mu\, {{r}_\star}-{2q\,Q\,{r}_\star^{-1} \Lambda_0}\\[3mm]\mu\, {{r}_\star}+{2q\,Q\,{r}_\star^{-1} \Lambda_0} &\kappa
\end{array}\right)
\left(\begin{array}{c}f \\g\end{array}\right)+O(1)\,,
\label
{truncated-}
\end{equation}
where $\Lambda_0^{-1}={2\sqrt{3{{r}_\star^2}+1}}$, and  $\kappa$ is an eigenvalue of the operator ${K}=-\beta\,({\boldsymbol{\tau}}\cdot{\bf L}+1)$.
We repeat, with due modifications, the procedure carried out in the previous section.
This gives ${\cal U}(r;E)$ whose leading term is 
\begin{equation}
-\frac{1}{({r}_\star^2-{r}^2)^2}\left[\frac{1}{4\Lambda^2_0}-\kappa^2
-{\mu^2{r}_\star^2}+\frac{{4}\left(qQ\right)^2}{ {r}_\star^2}\right].
\label
{remaining_singularity}
\end{equation}
This ${\cal U}(r;E)$ represents an infinitely deep potential well, Fig.~~\ref{eff-potential-two-horiz} (right).
With reference to the theorem about tunneling through a one-dimensional singular potential barrier \cite{Dittrich}, we state that a Dirac particle cannot escape from an extremal BH if and only if
\begin{equation}
{\mu^2{r}_\star^2}-\frac{{4}\left(qQ\right)^2}{ {r}_\star^2}\ge \frac34+\frac{1}{4\Lambda_0^2}-\kappa^2\,.
\label
{equilibr_condition}
\end{equation}
It is this additional condition that was mentioned above.

From Eq.~(\ref{equilibr_condition}) it follows that the Dirac particle is permanently confined to a cavity of radius $r={r}_\star$ if the combined effect of the attraction of the background gravitational and electromagnetic fields exceeds or is at least balanced by the centrifugal repulsion.
These effects are mixed in every term of Eq.~(\ref{equilibr_condition}) because any of them is present in ${r}_\star$ and $\Lambda_0$, which contain  both gravitational and electromagnetic contributions.
However, the very structure of these terms suggests that the main responsibility for the gravitational effect lies with the first term, and the main contribution of the electromagnetic effect comes from the second term.
The saturated form of Eq.~(\ref{equilibr_condition}) is 
\begin{equation}
\left(\mu{\cal M}\right)^2-\left({q}{\cal Q}\right)^2={\cal J}^2\,,
\label
{equilibr_satur}
\end{equation}
where  $\mu{\cal M}$ and ${q}{\cal Q}$ symbolize respectively the 
dominant  
gravitational and electromagnetic contributions to this relationship, and ${\cal J}^2$ is a positive (for not too large $|\kappa|$) quantity, symbolizing the centrifugal effect, or, equivalently, 
\begin{equation}
\left({\mu}{\cal M}-{q}{\cal Q}\right)\left({\mu}{\cal M}+{q}{\cal Q}\right)={\cal J}^2\,.
\label
{mathfrac_separ}
\end{equation}

Let ${q}{\cal Q}>0$.
This means that the electromagnetic field repels the Dirac particle from the center of the BH. Dividing both sides of equation (\ref{mathfrac_separ}) by the positive factor ${\mu}{\cal M}+{q}{\cal Q}$ gives
\begin{equation}
{\mu}{\cal M}={q}{\cal Q}+{C}\,,
\label
{mathfrac_eQ_positive}
\end{equation}
where  ${C}$ is a positive quantity associated with the centrifugal force which balances the forces of electromagnetic repulsion and gravitational attraction.
Equation (\ref{mathfrac_eQ_positive}) resembles the pseudospin symmetry condition (\ref{pseudospin}), if we take into account that the scalar field is equivalent to the symmetric  tensor field of the second rank in that both of them are such carriers of interaction that particles with charges of the same sign attract each other.

Let then ${q}{\cal Q}<0$.
This means that the electromagnetic field attracts the Dirac particle to the center of the BH.
Equation (\ref{mathfrac_separ}) can be reduced to
\begin{equation}
{\mu}{\cal M}=-{q}{\cal Q}+{C'}\,,
\label
{mathfrac_eQ_negative}
\end{equation}
where ${{C'}}$  is a positive constant similar in meaning to the constant ${C}$.  
Equation (\ref{mathfrac_eQ_negative}) is reminiscent of the spin symmetry condition (\ref{spin}).

We thus see that the dynamical affair of a Dirac particle as it proceeds along its orbit in the gravitational and electromagnetic fields of an extremal BH and is subject to the additional condition (\ref{equilibr_condition}) can be holographically mapped onto the dynamical affair of a quark driven by a mean field obeying the pseudospin symmetry condition (\ref{pseudospin}).

\section{Hunt for the lightest glueball ${\mathbb G}$}
\label
{glueball}
The idea of a bound system composed entirely of gluons arose in the early years of QCD development \cite{FritzschGell-Mann}--\cite{JaffeJohnson}.
Our concern here is with the lightest glueball, which will be designated by ${\mathbb G}$. 
This is a bound state of two gluons characterized by $J^{PC}=0^{++}$.
It has been established in \cite{Coleman} that there are no localized kink-type solutions with finite energy in pure Yang-Mills theory.
In other words, we have not the slightest notion what this system may have size and structure.
Unlike the quark, the gluon does not participate in the electromagnetic and weak interactions, and the same applies to ${\mathbb G}$. 
Since ${\mathbb G}$ is colorless and devoid of residual color interactions, it does not interact strongly with other hadrons untill it undergoes a phase transition, splitting into two separate gluons.
Lattice and sum rule calculations indicate that the mass of ${\mathbb G}$ is in the range from 1.3 to 2 GeV \cite{CloseT}--\cite{Crede}. 
Despite considerable efforts to detect ${\mathbb G}$, it has never been observed with complete certainty \cite{Crede}, \cite{Olive}. 
The widely held view is that the glueball field mixes with the quark-antiquark fields $\left(u{\bar u}+d{\bar d}\right)/\sqrt{2}$ and $s{\bar s}$ to form the observed meson resonances \cite{Ochs}.
To date, three isoscalar resonances with masses in the range from 1.3 to 2 GeV have been discovered: $f_0(1370)$, $f_0(1500)$, $f_0(1710)$ \cite{Olive}.

Thus, we are faced with twofold problem: (1) how to create glueballs in the unmixed state, and (2) how to identify the created  ${\mathbb G}$ by the characteristic products of its decay. 

\subsection{Photon colliders can create unmixed glueballs}
\label
{create_glueball}
\begin{figure}[htb]
\includegraphics[width=0.35\textwidth]{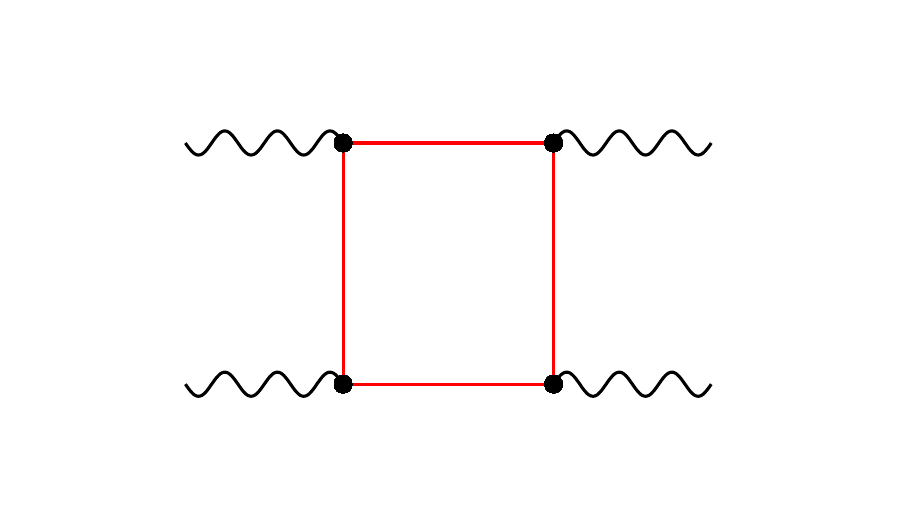}
\hskip-1.2cm{\includegraphics[width=0.35\textwidth]{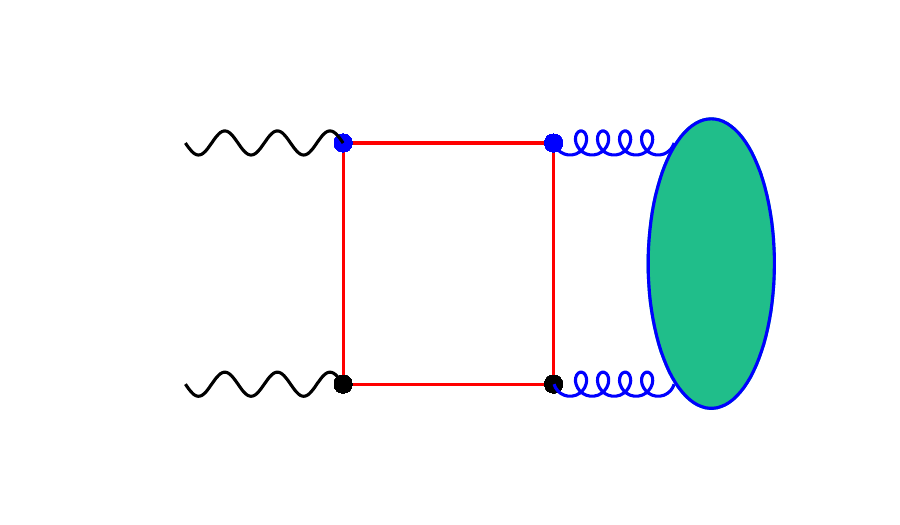}}
\hskip-1.2cm{\includegraphics[width=0.35\textwidth]{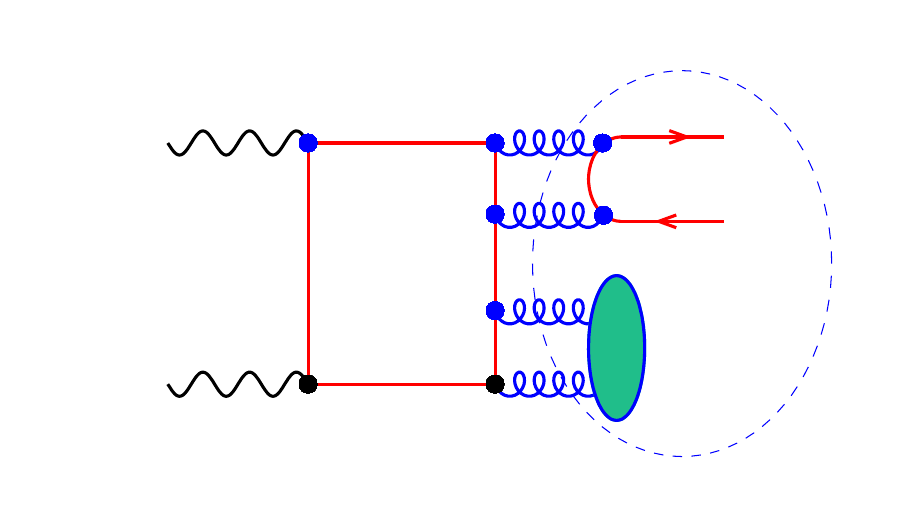}}
\begin{tabular}{lll}
\parbox{0.35\textwidth}{\centerline{(a)}}&
\parbox{0.25\textwidth}{\centerline{(b)}}&
\parbox{0.3\textwidth}{\centerline{(c)}}
\end{tabular}
\caption{$\gamma\gamma$ collisions produce the following: (a) two-photon system; (b) a glueball; (c) a merger of a glueball and a meson.
Photons, quarks, and gluons are shown, respectively, as sine waves, oriented straight lines, and spirals. 
The meson, composed of quark-antiquark pairs, is depicted as pairs of antiparallel rays 
}
\label{gamma-gamma}
\end{figure}

The very idea of preparing an unmixed scalar glueball can be taken from the pioneering works \cite{FritzschGell-Mann}--\cite{JaffeJohnson}.
They discussed  the decay of a scalar glueball into two photons with opposite polarizations.
By reversing this reaction, we arrive at the proposed in \cite{KPV-4} method for creating scalar glueballs.
Specifically, a head-on $\gamma\gamma$ collision with a center-of-mass energy $\sqrt s$ in the range from $1.3$ to $2$ GeV, provided that the helicity of the $\gamma\gamma$ system is zero, is expected to result in the creation of ${\mathbb G}$.
This reaction in the lowest order in $\alpha$ and $\alpha_s$ is represented by the Feynman diagram $(b)$ in Fig.~\ref{gamma-gamma}.   

\begin{figure}[htb]
\psfrag{LASERLIGHT}[c][c]{\tiny LASER LIGHT}
\psfrag{GQUANTA}[c][c]{\tiny $\gamma$-QUANTA}
\psfrag{ELECTRONBEAM}[c][c]{\tiny ELECTRON BEAM}
\psfrag{MAGNETICFIELD}[c][c]{\tiny MAGNETIC FIELD}
\psfrag{xx}[c][c]{${\bf x}_{\circ}$}
\centerline{\includegraphics[width=\risheight]{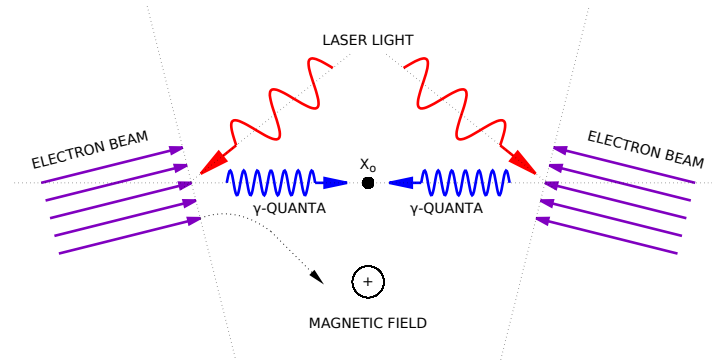}}
\caption{Photon collider}
\label{Telnov}
\end{figure}

To implement this plan, we need a {\it photon collider}.
This is a device in which laser photons are converted into high-energy $\gamma$-quanta due to their Compton scattering on high-energy electrons \cite{Ginzburg}.
The proposed experiment is set up as follows: laser photons are scattered by two beams of electrons moving towards each other to their collision point ${\bf x}_\circ$.
After scattering on the left electron beam, the laser photons are converted into $\gamma$-quanta which have an energy comparable to the energy of the electrons.
These $\gamma$-quanta move towards point ${\bf x}_\circ$, where they collide head-on with similar $\gamma$-quanta reflected from the electron beam on the right, as shown in Fig.~\ref{Telnov}.
The maximum energy $\omega$ of the $\gamma$-quanta obtained in this way is 
\begin{equation}
\omega\approx \frac{4E^2 \omega_0}{m^2+4E\omega_0}\,, 
\label
{omega-max}
\end{equation}
where $E$ and $\omega_0$ stand for the energy of the electrons and laser photons, respectively, and $m$ is 
the electron mass. 
For example, to convert photons with energy $\omega_0=1.17$ eV, emitted by a neodymium glass laser, into $\gamma$-quanta with energy $\omega=0.85$ GeV, electrons with energy of at lest  $E=7.5$ GeV is required. 
The energy spectrum of the $\gamma$-quanta is sharpest if the electrons are longitudinally polarized and the laser photons are circularly polarized. 
Among the existing electron accelerators, the Stanford Linear Accelerator could be most suitable as a key element of the carefully designed TESLA photon collider project \cite{B}.

To assess the feasibility of the $\gamma\gamma\to  {\mathbb G}$ conversion in the experimental setup under discussion, it is useful to turn to the experimental discovery of the scattering of light by light  
in quasi-real photon interactions of ultra-peripheral ${\rm Pb}-{\rm Pb}$ collisions, with impact parameters larger than twice the radius of the nuclei, at a center-of-mass energy $\sqrt{s}=5.02$ TeV  \cite{ATLAS_Collaboration}, \cite{Aad}.
Without going into detail of calculations of the two processes schematically represented by Feynman diagram $(a)$ and $(b)$ in Fig.~\ref{gamma-gamma} (the interested reader is referred to \cite{KPV-4}), we only mention the final conclusion: the creation of ${\mathbb G}$ in a head-on $\gamma\gamma$ collision with an energy $\sqrt s$ near $1.7$  GeV is characterized by a cross section 
$\sigma_{\gamma\gamma\to{\mathbb G}}\approx 60$ nb. 

An attractive feature of creating  ${\mathbb G}$ at a photon collider is that there are no mixing effects between the ground glueball state and the isoscalar $q{\bar q}$ mesons state.  
Indeed, from a comparison of diagrams $(c)$ and $(b)$ in Fig.~\ref{gamma-gamma} it is evident that the probability of the production of an aggregate of a glueball and a meson is suppressed by an overall factor of  $\alpha^4_s\sim 10^{-3}$ as against the production of an unmixed scalar glueball.

The expected luminocity at the TESLA photon collider of about $5\cdot 10^{33}$  ${\rm cm}^{-2}{\rm s}^{-1}$ \cite{B} is 7 orders of magnitude higher than the luminocity of $5\cdot 10^{26} {\rm cm}^{-2}{\rm s}^{-1}$ in the light-by-light scattering experiment in ${\rm Pb}-{\rm Pb}$ collisions  
\cite{ATLAS_Collaboration}, \cite{Aad}, which indicates the feasibility of a high-luminocity photon collider. 

\subsection{How to identify ${\mathbb G}$}
\label
{identify_glueball}
To examine the decay of ${\mathbb G}$ we invoke gauge/gravity duality because the lattice QCD is not yet able to analyze the interaction of glueballs with ordinary hadrons.
The new version of duality, discussed in Sec.~\ref{general holographic}, deals with the mapping of an extremal BH onto a {stable} subnuclear object, and hence does not seem to apply to the study of the ${\mathbb G}$ decay.
However, the most interesting thing is the dualism of {\it threshold situations}.
On the gravity side, this could be either a nearly extremal BH that has reached the final stage of Hawking radiation and is about to splits into two (or more) extremal BHs~\footnote{This splitting is a possible scenario for the end of the history of an evaporating BH. 
The scenarios proposed in the literature has been detailed in \cite{Chen}.}, or an extremal BH whose parameters approach the point of the naked singularity formation.
On the gauge side, threshold situations refer to either an unstable QCD system for which all types of decay modes are forbidden except one, or a stable system close to the stability failure.
We now turn to the former option~\footnote{The question may arise of whether such an almost stable particle is really dual to a nearly extremal BH, unlike an ordinary BH.
To verify this, we return to the two features of evaporating BHs that prevent duality (which were indicated at the beginning of subsection~\ref{novel_holographic}). 
First, if a nearly extremal BH, having exhausted its capacity for Hawking evaporation, splits into two extremal BHs, then this process is reversible, since the decay products can, in principle, collide and form a nearly extremal BH with zero Bekenstein temperature, an object incapable of Hawking evaporation. 
Second, if two Schwarzschild BHs had the same mass and Hawking evaporation rate at some instant, then at the end of their evaporation histories, the masses of these BHs will be equal, and so we are dealing with two identical objects.}. 

Since a phase transition  is required to split  ${\mathbb G}$ into two separate gluons, we assume that this phase transition is accompanied by breaking chiral ${\rm SU}(2)_L\times{\rm SU}(2)_R$ symmetry (characteristic of low-energy two-flavor QCD) down to isospin ${\rm SU}(2)_V$ symmetry.
Duality implies that the maximal spatial isometry ${\rm SO}(4)\sim{\rm SO}(3)\times{\rm SO}(3)$ of a non-rotating neutral nearly extremal BH is broken down to an ${\rm SO}(3)$ isometry when it splits into  two extremal BHs.
Such a  spatial isometry is possessed by a rotating BH if only one of its angular momenta is activated.
Again turning to duality, we conclude that ${\mathbb G}$ decays into two vector particles. 
Although gluons are vector particles, they cannot be decay products because they are color carrying objects. 
Therefore, the decay products of ${\mathbb G}$ are two {colorless vector particles}. 
In support of this implication we refer to the decay modes of two other neutral scalar particles,  $\pi^0$-meson and  Higgs boson $H$.
The decay of $\pi^0$ via the strong and weak channels is suppressed, so it decays as $\pi^0\to\gamma\gamma$, and $H$ does not participate in the strong and electromagnetic interactions, so it can decay only through the weak interaction channel, say into gauge vector bosons: $H\to W^+W^-$ and $H\to ZZ$.

\begin{figure}[htb]
\psfrag{G}[c][c]{${\mathbb G}$}
\phantom{.}
\hskip-3mm{\includegraphics[width=2.6cm,angle=270]{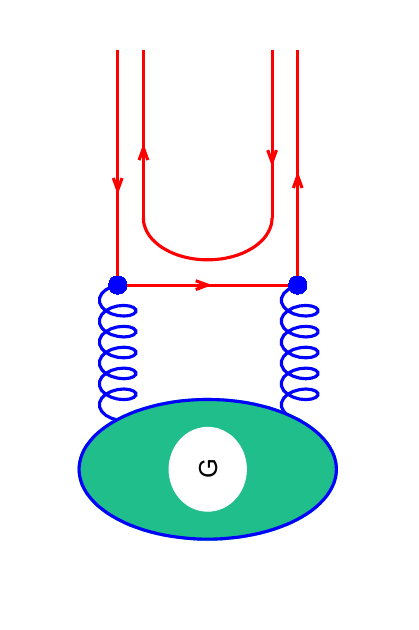}}
\hskip-5mm{\includegraphics[width=2.6cm,angle=270]{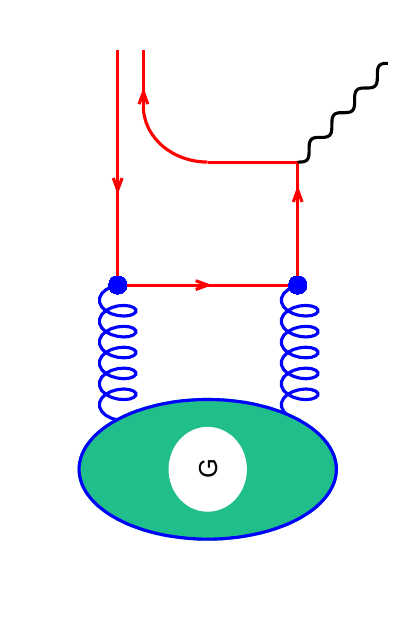}}
\hskip-5mm{\includegraphics[width=2.6cm,angle=270]{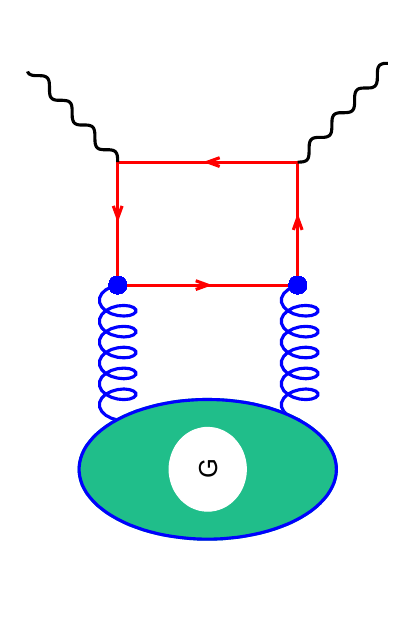}}
\hskip-5mm{\includegraphics[width=2.6cm,angle=270]{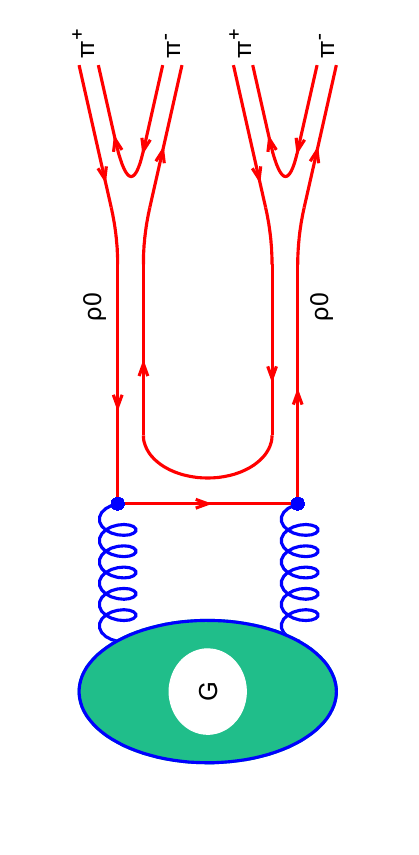}}\\
\begin{tabular}{cccc}
\parbox{0.25\textwidth}{\centerline{(a)}}&
\parbox{0.15\textwidth}{\centerline{(b)}}&
\parbox{0.2\textwidth}{\centerline{(c)}}&
\parbox{0.3\textwidth}{\centerline{(d)}}
\end{tabular}
\caption{Possible decay channels of ${\mathbb G}$}
\label{decay}
\end{figure}

The decay of ${\mathbb G}$ is represented by three Feynman diagrams (the lowest order in the coupling parameters $\alpha$ and $\alpha_s$ terms), Fig.~\ref{decay}. 
Diagram (a) shows a decay of ${\mathbb G}$ into $\rho^0\rho^0$, $m_{\rho^0}=775$ MeV, or
$\omega\omega$, $m_\omega=783$ MeV.
Diagram $(b)$ shows another decay mode into a photon and $\rho^0$ (or $\omega$ or $\phi$, $m_\phi=1019$ MeV).
Diagram $(c)$ displays a decay of ${\mathbb G}$ into two photons.
From the ratio of the probabilities of these decay channels, estimated at $1:{O}(\alpha):{O}(\alpha^2)$, we can state that the {main decay channel of ${\mathbb G}$ is ${\mathbb G}\to\rho^0\rho^0$}.
Since $\rho^0$ decays into $\pi^+\pi^-$ ($\approx 100${\%} fraction; $\Gamma=149.1\pm 0.8$ MeV \cite{Olive}), one may expect a significant increase in the yield of two $\pi^+\pi^-$ pairs, each of which has the angular momentum quantum number $l=1$, as $\sqrt s$ approaches to $m_{\mathbb G}$, Fig.~\ref{decay}, diagram $(d)$. 
This would clearly indicate of experimental detection of ${\mathbb G}$.

\section{Duality provides a way to find  $Z_{\rm max}$}
\label
{Holographic-}
Before calculating the maximum allowable electric charge $Z_{\rm max}$ in stable heavy nuclei, we need to clarify the rationale for the fact that the number of stable elements in the periodic table is limited~\footnote{Our concern here is only with {\it perfect} stability of nuclei.
We think of free protons as stable particles.}.
The mechanism for establishing the perfect stability fails in heavy nuclei when the wave functions of widely separated identical quarks no longer overlap~\footnote{The distances at which overlaps of the quark wave functions inside a heavy nucleus are still noticeable can be estimated by observing that the maximum allowable quark separation in stable nuclei of about 7 fm is found by applying Eq.~(\ref{R-nucleus}) to the ${}^{208}_{82}{\rm Pb}$ nucleus, while the magnitude of the wave functions of dressed $u$ and $d$ quarks becomes almost negligible at the distance of their Compton wave lengths, $\lambdabar_{u,d}\approx \frac23$ fm.}, which releases these quarks from control of the Pauli exclusion principle.
Accordingly, the degeneracy pressure does not increase  in proportion to the number of identical quarks, and the equilibrium between attraction and repulsion no longer holds.
The failure of validity of the Pauli exclusion principle means that we have reached the boundary
beyond which the quantum laws cease to apply, giving way to the classical laws. 

\subsection{Quantum vs classical}
\label{quantumclassical}
What is the difference between classical and quantum objects?
Classical reality assigns a clear ontological status to any material entity, say, a particle either exists or does not, {there is no third option}. 
However, the structure of the physical world is not required to exhibit existential certainty.
The existence of its structural elements may be {random}.
This is exactly what happens in quantum reality.
Its elements are still called particles, but from the point of view of a classical observer, the behavior of such particles obeys {probabilistic} laws.
Some particles can {spontaneously} convert into other particles.
For example, a free neutron is prone to  $\beta$-decay,
\begin{equation}
n\to p+ e +{\bar \nu}_e\,.
\label
{d_to_u_e_nu}
\end{equation}  
The state of a neutron is a superposition of two states: the state of the undecayed neutron and the state of its decay products.
Therefore, at any moment there is some probability $W$ that this neutron still exists, but with the probability of $1-W$ it no longer exists.
The absence of existential certainty is also typical for stable quantum objects such as electrons.
Unlike a classical particle, which retains its individuality throughout its history, an electron does not care about preserving its identity; instead, it exhibits identity with and is indistinguishability from any electron in the universe,  including an electron that has not yet been produced in  reaction (\ref{d_to_u_e_nu}).
Any case of isolation and individualization of an electron among an infinite number of identical and indistinguishable electrons (quanta of the Dirac electron field) can be regarded as a violation of quantum order and the appearance of a fragment of the classical nature in the quantum picture~\footnote{In the currently accepted terminology, such cases are referrered to the processes of ``observing'' an electron, or ``measuring'' its properties, however, more generally, one should speak of the intrusion of the existentially certain (classical) structure of reality into the immanently random (quantum) structure.}.

The Pauli exclusion principle is one of the cornerstones of quantum physics.
Therefore, the end of nuclear stability due to the loss of control of the Pauli exclusion principle over the behavior of identical quarks in heavy nuclei can be associated with the penetration through the demarcation line between the quantum and the classical.
The search for the exact location of this demarcation line could be greatly facilitated if we could first find its pictorial rendition.
This is a clear hint that to find $Z_{\rm max}$ we should turn to gauge/gravity duality.
Indeed, it is possible to show \cite{K-2008} that the {event horizon of a BH is an interface between the classical and the quantum}~\footnote{The essence of the matter is simple. 
In a spherically symmetric geometry, the interchange of  $t$ and  $r$ entails the replacement of the retarded Green's function ${\tilde D}_{\rm ret}(p)$, responsible for the propagation of a classical free field $\Phi$, by the St\"uckelberg--Feynman propagator ${\tilde D}_c(p)$, which ensures the propagation of the corresponding quantized field  ${\hat \Phi}$.
On the other hand, the replacement of the metric signature, $(+1,-1,-1,-1)\Longrightarrow(-1,+1,+1,+1)$, occurs when passing from the outer region of a BH to its inner region. 
In other words, the laws of propagation for classical and quantized fields swap around when the event horizon of a BH is crossed. 
The event horizon can thus be viewed as an interface between the classical and  the quantum.
The interested reader may take a closer look at this phenomenon in \cite{K-2008}.}.
If the event horizon disappears and  a {naked singularity is formed} in a  BH located in ${\rm AdS}_5$, violating  the quantum-classical arrangement, this is supposedly mapped onto the ruin of nuclear stability in ${\mathbb R}_{1,3}$ \cite{KPV-5}.

\subsection{The occurrence of a naked singularity is dual to the end of nuclear stability}
\label
{Holographic}
The general solution of the Einstein--Maxwell--Chern--Simons set of equations describes the gravitational and electromagnetic fields of a charged rotating BH in AdS${}_5$   \cite{Cvetic}.
The radial part of the metric takes the form:
\begin{equation}
g_{rr}=r^{2}\left({r^2+a^2\cos^2\theta+b^2\sin^2\theta}\right){\Delta_r}^{-1}\,,
\label{metric}
\end{equation}
\begin{equation}
{\Delta_r}=\left(r^2+a^2\right)\left(r^2+b^2\right)\left({r^2}+1\right)-2Mr^2
+{(Q+ab)^2-a^2b^2}\,,
\label{Delta-r}
\end{equation}
where  $M, Q, a, b$ are, respectively, the mass, charge, and two angular momenta of the BH.
From the extremality condition, 
\begin{equation}
r^6+{\mathfrak a}r^4+{\mathfrak b}r^4+(Q+ab)^2=0,
\label{Delta=0}
\end{equation}
 where
\begin{equation}
{\mathfrak a}=a^2+b^2+1\,,
\quad
{\mathfrak b}=a^2+b^2-2M+a^2b^2\,,
\label
{A-df}
\end{equation}
we obtain the expressions for the event horizon radius $r_0$ and the charge $Q$ of the BH,
\begin{equation}
3r_0^2=\sqrt{{\mathfrak a}^2-3{\mathfrak b}} -{\mathfrak a}\,,
\qquad
Q=r_0^2\sqrt{2r_0^2+{\mathfrak a}}-ab\,.
\label
{root-}
\end{equation}

The Dirac equation  with the Pauli term for the anomalous magnetic moment 
\[
\left[\gamma^A e^\alpha_A\left(\partial_\alpha+\Gamma_\alpha-iqA_\alpha\right)+\frac{i}{4\sqrt{3}}\gamma^\mu\gamma^\nu F_{\mu\nu}+\mu\right]\chi(x)=0\, 
\]
allows separation of variables \cite{Wu}. 
To simplify the analysis, we focus on the radial part of the Dirac equation near the {event horizon}, 
\begin{equation}
\frac{d}{dr}
\left(\begin{array}{c}
f \\g\end{array}\right)
=
\frac{1}{r-r_0}
\left(\begin{array}{cc}
-\dfrac12+\dfrac{A_0}{\sqrt{c_0}} & \dfrac{D_0}{{c_0}}-\dfrac{B_0}{\sqrt{c_0}}\\[3mm]
-\dfrac{D_0}{{c_0}}-\dfrac{B_0}{\sqrt{c_0}} &-\dfrac12-\dfrac{A_0}{\sqrt{c_0}}
\end{array}\right)
\left(\begin{array}{c}f \\g\end{array}\right)
+O(1)\,,
\label
{truncated}
\end{equation}
\[
A_0=\lambda r_0-\frac{Q+ab}{2r_0}\,,\,\,\,
B_0=\mu r_0^2-abE+mb\left(1-a^2\right)+ka\left(1-b^2\right)\!,\,\,\,
c_0=4r_0^2\left(3r_0^2+{\mathfrak a}\right)\!,
\]
\[
D_0=4Er_0^3+2r_0 \left[E\left({\mathfrak a}-1\right)-ma\left(1-a^2\right)-kb\left(1-b^2\right)-
\frac{\sqrt{3}}{2}qQ\right].
\]
Here, $\lambda$ is the separation constant, and $m$ and $k$ are the magnetic quantum numbers  associated with the two independent angular momenta $a$ and $b$ of the BH.
The system of two first-order differential equations (\ref{truncated})  reduces to a  one-dimensional second-order Schr\"odinger-like equation 
\begin{equation}
F''+\frac{{\cal U}_0}{\left({r}-{r}_{0}\right)^{2}}\,F=0\,,
\label
{Schr-like}
\end{equation}
where
\begin{equation}
{\cal U}_0=
\left(\frac{D_0}{c_0}\right)^2+\frac14 -\frac{A_0^2+B_0^2}{c_0}\,.
\label
{U-0}
\end{equation}
The confinement of the Dirac particle is lost if, in the limit $r_0\to 0$, the event horizon disappears
and a naked singularity forms.
Note that $c_0\sim 4r_0^2{\mathfrak a}$, which renders ${\cal U}_0$ divergent as $r_0\to 0$.
We require that the divergent contributions in $\left(D_0/c_0\right)^2$ and $B_0^2/c_0$ cancel.
This leads to a set of algebraic equations~\footnote{For an extended description of the computational procedure, see \cite{KPV-5}.}.
Solving them yields ${\mathfrak a}\approx 2.58$.
The set of equations  is then reduced to a single equation 
\begin{equation}
qQ=\frac{1}{8\sqrt{3}}\left({\mathfrak a}-1\right)^2\left(2-\sqrt{{\mathfrak a}}\right)
\left({\mathfrak a}+3\right).
\label
{qQ}
\end{equation}
Substituting ${\mathfrak a}\approx 2.58$ into (\ref{qQ}) yields $qQ\approx 0.396$. 
The projection of $qQ$ onto ${\mathbb R}_{1,3}$ is the charge of the $u$ quark multiplied by the nuclear charge $Z_{\max}$, containing this $u$ quark, minus the charge of the quark, 
\begin{equation}
qQ=\frac23\left(Z_{\max}-\frac23\right)\alpha\,,
\label
{final-Z}
\end{equation}
where $\alpha$ is the fine-structure constant,  $\alpha\approx 1/137$.
Ultimately,
\begin{equation}
Z_{\max}\approx \frac23+137\,\,
\frac32\,\, 
0.396\approx 82\,.
\label
{final-Z-}
\end{equation}
The calculated $Z_{\max}$ equals the electric charge of the ${}^{208}_{82}{\rm Pb}$ nucleus \cite{KPV-5}.

\section{Summary and outlook}
\label
{Conclusion}
 \begin{itemize}
  \item
Quark-based models allow us to study the {composition and static properties of
stable nuclei} and to understand why these  properties change as $Z$ increases.
  \item
The Fermi gas model explains why, in light stable nuclei up to ${}^{40}_{20}{\rm Ca}$, {the numbers of $d$ and $u$ quarks are nearly equal},  $n_u\approx n_d$. 
In particular, this model explains why {stable systems composed solely of neutrons} are  {impossible}.
  \item
The modified bag model accounts for the quark composition of stable nuclei heavier than  ${}^{40}_{20}{\rm Ca}$.  
Calculations of nuclear magnetic dipole moments within this framework agree with experimental data to within $\sim 10\%$ (with a few $\sim 20\%$ outliers) for a large set of stable isotopes.
  \item
Using a new version of duality, it is possible to predict the {dominant decay channel of the lightest glueball},  ${\mathbb G}\to\rho^0\rho^0\to\pi^+\pi^-+\pi^+\pi^-$.
  \item
Duality provides a way to {find $Z_{\max}\approx 82$}, which is precisely the charge of  ${}^{208}_{82}{\rm Pb}$ .
\end{itemize}

In the approach developed in Refs.~\cite{KPV-1}--\cite{KPV-3} and \cite{KPV-5} the emphasis was on the quark composition and static properties of {\it stable} nuclei.
This limitation of the subject reveals both the strength and weakness of the exploited capabilities of  quark-based models.
We are presented with a clear and physically sound understanding of the balance of forces, that is, the equilibrium between the degeneracy pressure associated with the Pauli exclusion principle and the infrared QCD effect of mutual attraction of widely separated quarks. 
This made it possible to roughly reproduce the entire nuclear stability archipelago depicted in Fig.~\ref{stable}, in particular to display with reasonable accuracy part of the experimentally established curve of the average binding energy per nucleon for $A\ge 40$, cf. the left and right drawings in  Fig.~\ref{comparison}.
By applying the powerful technique of algebraic geometry to study of the cubic curve,  Eq.~(\ref{Delta=0}), which determines the extremality condition of BHs in ${\rm AdS_5}$, it is very likely to give an accurate description of the  nuclear stability archipelago in all details. 
For example, this could open a new avenue of attack on the problem of lithium isotope abundance: why do the stable isotopes ${}^6_3{\rm Li}$ and ${}^7_3{\rm Li}$ occur in nature in a ratio of 7.5\% and 92.5\%?
To what extent is this fact due to primordial and stellar nucleosynthesis, and to what extent to the peculiarities of the quark structure of these nuclei?
On the other hand, in models of nuclei developed within the Yukawa paradigm, long-range attractive forces are absent, and hence a convincing explanation of the perfect stability of heavy nuclei is highly improbable. 
Furthermore, the empirical fact that spontaneous emission of pions by nuclei is absent suggests that pions are alien to nuclear physics as adequate building blocks in low-energy effective theories to QCD~\footnote{Of course, this does not mean that there is a gap between the quark-based models and the models following the Yukawa paradigm.
Indeed, the essence of what is suggested in \cite{KPV-1}--\cite{KPV-3} and \cite{KPV-5} is simply the adoption of quark variables as primary, fundamental degrees of freedom in nuclear physics, while nucleons and pions may, where physically appropriate,  be derived degrees of freedom associated with collective effects, similar to quasiparticles in condensed matter physics.}.

An important open issue remains how to extend the analysis developed in \cite{KPV-1}--\cite{KPV-3} and \cite{KPV-5} to nuclei that are not perfectly stable but have a long -lifetime.
One can try to weaken the requirement for the singularity of the effective potential ${U}\left(r;\varepsilon\right)$, and allow a  singularity $\sim \gamma\left(r-r_\ast\right)^{-2}$ with $0<\gamma<\frac34$.
This would make it possible to calculate the probability of a tunnel transition through such  potential barriers.
However, we are not concerned with the tunneling of a single quark (which is physically infeasible) but with the tunneling of colorless clusters, such as neutrons or alpha particles, which is much more difficult task.
Moreover, it is not entirely clear whether it is possible to find a gauge/gravity version of duality between a nucleus in ${\mathbb R}_{1,3}$ with such a penetrable effective potential  ${U}\left(r;\varepsilon\right)$ and its black hole counterpart in AdS${}_5$.
In this context, the intensively developed black hole chemistry \cite{Mann} with all kinds of phase transitions in  AdS${}_5$ and their holographic mapping on ${\mathbb R}_{1,3}$ seems very promising.  

A major step in refining quark-based nuclear models could come from the general progress in further development of QCD, especially the creation of a complete description of the  quantum self-interacting quark taking into account various types of environment~\footnote{It seems exceptionally intriguing to speculate that a QCD fluctuation could lead to such a heavy dressing of $d$ quarks in a nucleus at the edge of the stability state as to allow beta decay of one of these quarks, $d\to u+ e +{\bar \nu}_e$, and thus relegate this stable nucleus to the status of a merely long-lived system.}
 and the construction of a theory of scattering of hadrons on nuclei and nuclei on nuclei, say, in the spirit of the Harari--Rosner dual diagram technique \cite{Harari}, \cite{Rosner}.

\section*{Acknowledgments}
We thank Yaroslav Alekseev, Stanley Brodsky, Masud Chaichian, Evgeny Epelbaum, Joseph Ginocchio, Wolfgang Ochs, Douglas Singleton, Valery Telnov, Vicente Vento, Richard Woodard, and Koichi Yazaki for fruitful discussions over the past decade as we worked on the problems raised in Refs.~\cite{KPV-1}--\cite{KPV-5}.
We are especially grateful to John Klauder and Terry Goldman for their enthusiastic reception of  our analytical and conceptual  findings.

\section*{Conflict of interest statement}
The authors declare that they have no competing interests.

\section*{Data access statement}
No data supporting the research are  included in this manuscript.

\section*{Funding statement}
This work was funded from the budget of the Russian Federal Nuclear Center--VNIIEF. 
No additional grants or sources of funding were received for the conduct or supervision of this study.

\end{document}